\documentclass[sigconf]{acmart}
\AtBeginDocument{%
  }

\usepackage{subcaption}

\acmISBN{978-1-4503-XXXX-X/18/06}

\usepackage{booktabs}
\usepackage{multirow}
\begin{document}

\title{Two Birds with One Stone: Improving Rumor Detection by Addressing the Unfairness Issue}

\author{Junyi Chen}
\affiliation{%
  \institution{University of Auckland}
  \country{New Zealand}}

\author{Mengjia Wu}
\affiliation{%
  \institution{University of Technology Sydney}
  \country{Australia}
  }
\author{Qian Liu}
\affiliation{%
  \institution{University of Auckland}
  \country{New Zealand}}
\author{Ying Ding}
\affiliation{%
  \institution{University of Texas at Austin}
  \country{USA}}
\author{Yi Zhang}
\affiliation{%
  \institution{University of Technology Sydney}
  \country{Australia}}


\begin{abstract}
The degraded performance and group unfairness  caused by confounding sensitive attributes in rumor detection remains relatively unexplored. To address this, we propose a two-step framework. Initially, it identifies confounding sensitive attributes that limit rumor detection performance and cause unfairness across groups. Subsequently, we aim to learn equally informative representations through invariant learning. Our method considers diverse sets of groups without sensitive attribute annotations. Experiments show our method easily integrates with existing rumor detectors, significantly improving both their detection performance and fairness.
\end{abstract}





\maketitle
\section{Introduction}
Most existing rumor detectors \cite{devlin2018bert,nan2021mdfend,wang2018eann,azri2021calling} utilize data-driven methods to learn news content representations for rumor classification. This follows an ideal causal pathway $x \rightarrow y$, where $x$ denotes news content and $y$ represents the classification label. However, they often neglect the unseen confounding effect modeled as $x \leftarrow s \rightarrow y$, where a sensitive attribute\footnote{We loosely refer to sensitive attributes as those that define groups relevant to the benefits of the rumor detector's outcomes.} $s$ indicates group membership (e.g., domain) of the content. Taking domain-based groups as an example, $x \leftarrow s$ manifests as domain-specific linguistic patterns shape the content (e.g., science versus politics). The path $s \rightarrow y$ manifests through data collection biases, demonstrated by the disparate class distribution across domains \cite{zhou2024finefake}. This extends to other sensitive attributes (e.g., platform), where news from major agencies differs linguistically from social media streams and is stereotyped as more trustworthy.
Hence, non-causal shortcuts can be learned given the confounding effect, degrading detection performance and fairness.
Recent works \cite{zhang2022causalrd,zhu2022generalizing,chen2023causal} only note dataset bias but still solely pursue detection performance gains, largely neglecting unfairness for different sets of groups. We reveal two critical challenges that impede fair, high-performing rumor detection: \textbf{1. Limited annotated but diverse sensitive attributes.} Privacy constraints prevent comprehensive sensitive attribute collection  (e.g., domains, platforms, authors' certification status, political leaning, etc.), yet each attribute defines its own set of groups. \textbf{2. Variant feature learning.} Limited sensitive attribute supervision produces features that remain unfairly variant to unknown sets of groups, compromising generalization of fairness and detection effectiveness.

This work studies rumor detection from an underexplored fairness perspective. We present a max-min framework that can be plugged into existing rumor detectors to promote fairness while improving detection performance without requiring annotated sensitive attribute data. In the first step, we partition the training data into different subsets that lead to the worst-case performance of both detection and fairness for a rumor detector. This simulates the  sensitive attributes that can cause non-causal and unfair predictions across a set of groups. In the second step, we aim to improve and balance the performance across each subset (group) created by the partitions, thereby learning invariant feature representations for rumor detection. 
Since the partitioning operation does not rely on annotated sensitive attributes, it results in dynamic partitions during each training epoch alongside parameter updates, allowing for the consideration of extensive sets of groups. 
This ensures that the learned invariant feature representations generalize across different sets of groups. Note that this study does not aim to challenge the usefulness of auxiliary (potentially sensitive) but emphasizes preventing  models from unfairly over-relying on those confounding sensitive attributes when predicting.
Our key contributions are:
\begin{itemize}
\vspace{-1.3mm}
     \item We study rumor detection from a fairness perspective regarding multiple sets of groups, a novel angle largely overlooked in previous research.
    \item We propose a unified fairness framework that considers various sets of groups without requiring sensitive attribute annotations.
    \item Convincing experiments demonstrate that our method significantly improves both detection performance and fairness of existing rumor detectors across multiple sets of groups in a plug-and-play manner.
\end{itemize}
\begin{figure}
    \centering
    \includegraphics[width=\linewidth]{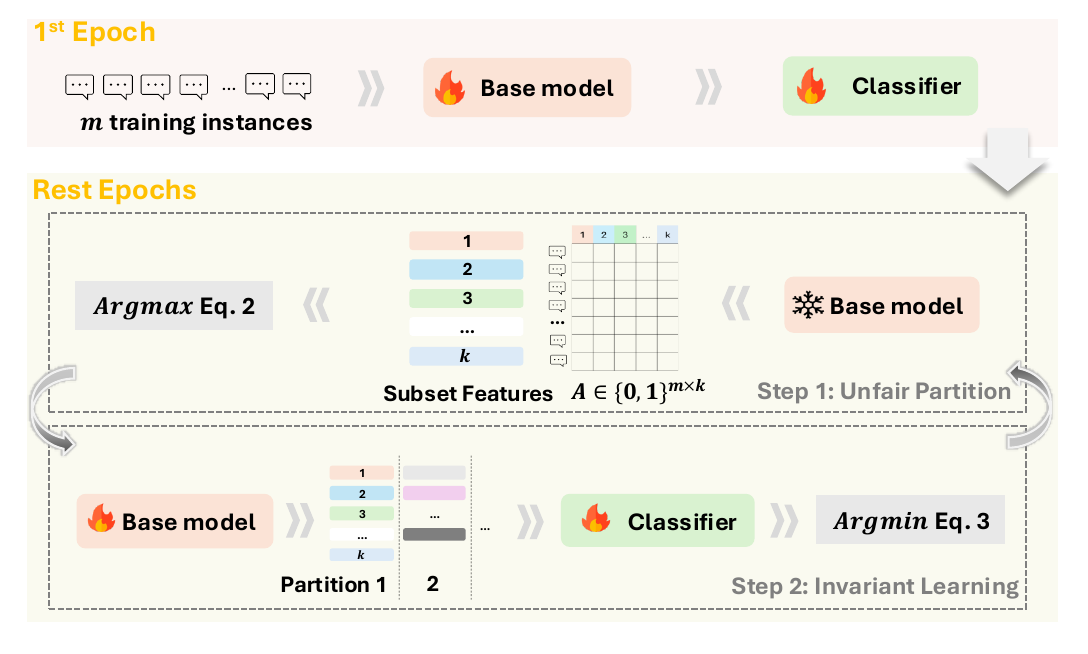}
    \caption{Schematic view of our proposed method.}
    \label{fig:framework}
\vspace{-0.4cm}
\end{figure}
\section{Method}
In Fig. \ref{fig:framework}, our method begins with standard training and then repeatedly performs two steps: (1) Unfair Partition and (2) Invariant Learning.

\noindent\textbf{Problem Formulation.} Let $c = \{\boldsymbol{x}, y, \boldsymbol{S}\}$ represent a rumor detection instance, where $\boldsymbol{x}$ denotes the initial embedded feature, $y$ the classification label, and $\boldsymbol{S} = \{\boldsymbol{s}_1, \ldots, \boldsymbol{s}_n\}$ the sensitive attribute vectors indicating group membership of different sets (e.g., $\boldsymbol{s}_1$ indicates the domain, $\boldsymbol{s}_2$ indicates the platform). Given a rumor detector with a base model $\Phi(\cdot)$ for feature extraction and a classifier $f(\cdot)$, we aim to optimize both for competitive binary prediction $y \in \{\text{rumor}, \text{non-rumor}\}$ while ensuring fairness across groups from all potential sets. $S$ serves only for fairness evaluation. We omit the subscript $i$ when the context is clear.
\subsection{Unfair Partition}
The concept of backdoor adjustment \cite{pearl2016causal} suggests statically partitioning training data to subsets based on sensitive attributes to learn fair representations. However, this approach demands comprehensive attribute annotations - impractical for social media data and risks overlooking unidentified sets of groups.

Inspired by invariant learning \cite{li2024learning,zhu2024one}, we propose  dynamic data partitioning to consider multiple sets of groups during training, without requiring sensitive attribute annotations. Our method identifies "unfair partitions" where the frozen base model performs worst.  Recalling the ideal causal pathway $\boldsymbol{x} \rightarrow y$, if the base model learns to extract causal features and is unaffected by confounding sensitive attributes, it should generally produce well-separated features based on contrastive labels $y$ within each subset given a partition. Otherwise, it may be influenced by confounding attributes via the pathway $\boldsymbol{x} \leftarrow s \rightarrow y$, where the confounding effect of $s$ varies across instances within and between subsets, leading to non-causal, non-generalizable features and thus, the worst performance within a partition. In this context, subsets in the partition can represent a set of groups conditioned on the specific type of confounding sensitive attribute (e.g., content from the domain of science versus politics).

Based on the discussion above, for an instance $c$ with its initial embedded feature $\boldsymbol{x}$, we first deduce the partition:
\begin{equation}
    \boldsymbol{\hat{s}} = FC(\Phi(\boldsymbol{x})),
\end{equation}
where $FC(\cdot)$ is a fully-connected layer with trainable parameters and $\boldsymbol{\hat{s}} \in \mathbb{R}^{1 \times k}$ indicates which subset the instance belongs to. 

By applying this operation across $m$ instances, we construct a partition matrix $\boldsymbol{A} \in \{0,1\}^{m \times k}$, which we want to potentially reflect data distribution under a specific set of groups. Based on the objective of invariant risk minimization  \cite{arjovsky2019invariant}, we identify unfair partitions where the base model performs worst, helping reveal confounding sensitive attributes:
\begin{equation}
    \underset{\boldsymbol{A}}{\operatorname{argmax}} \sum_k \mathcal{L}_\mathrm{con}(\boldsymbol{A}, k, \Phi (\boldsymbol{x}), y) + \lambda \operatorname{Var}\left(\left\{\mathcal{L}_\mathrm{con}(\cdot)\right\}_{i=1}^k\right),
\end{equation}
where $\mathcal{L}_\mathrm{con}$ denotes the supervised contrastive loss for each subset given a partition, $\operatorname{Var}$ measures performance variance across subsets and $\lambda$ is a trade-off factor. In implementation, we treat the items within the same subset that share the same label $y$ as positive pairs, while the remaining items within the same subset are treated as negative pairs.
A higher value in the first term indicates the base model's reliance on a potential confounding sensitive attribute - a non-causal correlation that limits detection performance. A higher value in the second term indicates performance disparity across subsets. Using supervised contrastive loss directly help us measure the feature variance and thus reveal potential confounding sensitive attributes.
\begin{table*}[t]
\centering
\caption{Performance comparison of base models with our method, formatted as `original $\pm$ improvement'.}
\label{tab:performance}
\begin{tabular}{@{}llllllllll@{}}
\toprule
\multirow{2}{*}{Base Model} & \multicolumn{5}{c}{\textit{FineFake}}                                                 & \multicolumn{4}{c}{\textit{PHEME}}                 \\ \cmidrule(l){2-10} 
\multicolumn{1}{c}{}                            & Acc  $\uparrow$     & F1 $\uparrow$      & $\Delta_{d}$  $\downarrow$       & $\Delta_{p}$ $\downarrow$         & \multicolumn{1}{l|}{ $\Delta_{a}$ $\downarrow$ }         & Acc  $\uparrow$  & F1  $\uparrow$    & $\Delta_{d}$ $\downarrow$       & $\Delta_{a}$ $\downarrow$       \\ \cline{1-10} 
BERT                                            & 72.6\textbf{+3.1} & 71.9\textbf{+2.3} & 41.9\textbf{-12.4}  & 86.9\textbf{-7.1}  & \multicolumn{1}{l|}{36.0\textbf{-7.5}}  & 65.3\textbf{+2.1} & 64.1\textbf{+3.1} & 80.8\textbf{-14.4} & 40.2\textbf{-8.1} \\
DualEmo                                         & 74.4\textbf{+3.1} & 73,2\textbf{+3.0} & 35.5\textbf{-5.0} & 94.4\textbf{-5.4}  & \multicolumn{1}{l|}{42.9\textbf{-2.9}}  & 73.2\textbf{+2.1} & 73.0\textbf{+2.0} & 85.2\textbf{-10.3} & 44.6\textbf{-3.7} \\
EANN                                            & 75.5\textbf{+3.2} & 75.8\textbf{+2.2} & 34.7\textbf{+1.2}  & 89.8\textbf{-11.8} & \multicolumn{1}{l|}{39.8\textbf{-7.2}}  & 72.2\textbf{+3.1} & 71.5\textbf{+4.0} & 80.3\textbf{-12.2} & 48.2\textbf{-7.9} \\
MDFEND                                          & 76.2+\textbf{2.1} & 76.1\textbf{+1.5} & 40.3\textbf{-10.1} & 98.4\textbf{-18.4} & \multicolumn{1}{l|}{44.6\textbf{-11.0}} & 78.3\textbf{+7.9} & 78.2\textbf{+7.5} & 83.7\textbf{-12.1}& 51.3\textbf{-8.6}\\
\bottomrule
\end{tabular}
\end{table*}

At this stage, we freeze the base model.
Each epoch's partition matrix $\boldsymbol{A}$ is added to set $\mathbb{A}$, enabling continuous discovering of unfair partitions caused by different confounding attributes and thus, different sets of groups.  We maintain only the six partition matrices with the highest loss, based on empirical findings. 

\subsection{Invariant Learning}
The invariant learning step eliminates non-causal correlations induced by confounding sensitive attributes within each partition. On one hand, it drives both the base model and classifier to improve detection performance, and on the other hand, it promotes balanced performance across groups for $\forall \boldsymbol{A} \in \mathbb{A}$, to learn more invariant features. Hence, we flip the objective to minimize it:
\begin{equation}
    \underset{\Phi, f}{\operatorname{argmin}} \sum_k \mathcal{L}_\mathrm{cls}(\boldsymbol{A},k,f, \Phi(\boldsymbol{x}),y) + \lambda \operatorname{Var}\left(\left\{\mathcal{L}_\mathrm{cls}(\cdot)\right\}_{i=1}^k\right), \quad \forall \boldsymbol{A} \in \mathbb{A}
\end{equation}
where $\mathcal{L}_\mathrm{cls}$ is the cross-entropy loss used for classification.
\section{Experimental Evaluation}
\subsection{Setup}
\noindent\textbf{Datasets.} We conduct experiments on two  benchmarks: (1) \textit{FineFake} \cite{zhou2024finefake} (10,507 non-rumors, 6,402 rumors) with three types of sensitive attributes: 8 platforms, 6 domains, and author certification status; and (2) \textit{PHEME} \cite{zubiaga2017exploiting} (1,428 non-rumors, 590 rumors) with two types of sensitive attributes: 6 event domains and author certification status. Both datasets have removed duplicate entries.

\noindent\textbf{Base Models.} Our proposed method can be applied to existing rumor detection base models in a plug-and-play manner. We evaluate its effectiveness on several popular models: (1) purely content-based method BERT \cite{devlin2018bert}, (2) emotion-based method DualEmo \cite{zhang2021mining}, and (3) multi-domain-based methods EANN \cite{wang2018eann} and MDFEND \cite{nan2021mdfend}. Only source text content is used in our experiments.

\noindent\textbf{Evaluation Protocol.} We split the data 6:1:3 into train/validation/test sets randomly. Methods are trained for 200 epochs with early stopping after 10 epochs without validation improvement. The best checkpoint is used for testing. For detection performance metrics, we use Accuracy and F1. For fairness metrics, $\Delta_{d/p/a}$ denotes the maximum Demographic Parity \cite{barocas2023fairness} across topic (event) domains, platforms, and authors. We set $k=2$ and $\lambda=0.1$ for all experiments and omit hyperparameter analysis due to page constraints.
\subsection{Result}
\subsubsection{Performance Comparison}
We present the results in Table~\ref{tab:performance}, which reveals several key findings. (1) Our proposed method improves both fairness metrics and detection performance across all base models, confirming that domains, platforms, and author certification status act as confounding factors. This also verifies that our method effectively handles multiple sets of groups.
(2) Improvements on BERT and DualEmo suggest that  language models can easily learn linguistic and emotional shortcuts, which worsen detection performance and exacerbate unfairness.
(3) For EANN, we maintain competitive or even superior fairness metrics on the domain side while significantly improving the other two. We attribute this to EANN's use of adversarial training to remove domain information from the final learned feature, which effectively promotes fairness in the domain aspect but falls short in other areas, where we observe the fairness on platforms and authors is worse than naive baseline BERT. This endorses our assumption that focusing solely on a specific set of group may result in feature representations that are more detrimental to unconsidered sets of groups. Note that EANN requires domain annotations, whereas our method does not. (4) MDFEND aims to learn domain-specific features to support final detection. While its detection performance is impressive, it notably exacerbates the unfairness issue, even in the domain aspect. This is understandable, as aggressively utilizing domain features without considering fairness can easily lead to stereotypes that improve overall detection performance but compromise fairness across domains and other dimensions. Our method addresses this issue by enforcing more generalizable invariant features that improve both performance and fairness, particularly on \textit{PHEME}. This ensures with our method, the base model can effectively leverage auxiliary attributes without over-relying on them.
\subsubsection{Ablative Study}
We fully justify our design motivations via following variants:
(1) w/ SP: using static partition on training data with all ground truth sensitive attribute labels instead of our dynamic partition strategy; 
(2) w/o $\mathcal{L}_\mathrm{sup}$: replacing the supervised contrastive loss with typical cross-entropy loss in the unfair partition step; 
(3) w/o Var: dropping the variance loss between partitioned subsets for both stages; and
(4) w/o Record: not recording every partition during training, instead using only the latest one.
Due to page limitations, we demonstrate only the results using the base model BERT as the observations remain largely consistent across other models.
\begin{table}[t]
\centering
\caption{Ablative study on model variants.}
\begin{tabular}{@{}lllll@{}}
\toprule
\multirow{2}{*}{Variants} & \multicolumn{2}{c}{\textit{FineFake}}               & \multicolumn{2}{c}{\textit{PHEME}} \\ \cmidrule(l){2-5} 
                          & Acc  & \multicolumn{1}{l|}{$\Delta_{d/p/a}$}          & Acc       & $\Delta_{d/a}$           \\ \midrule
BERT w/ ours              & 75.7 & \multicolumn{1}{l|}{29.5/79.8/28.5} & 67.4      & 66.4/32.1     \\ \midrule
w/  SP                    & 74.2 & \multicolumn{1}{l|}{28.4/80.1/27.9} & 65.9      & 69.2/30.8     \\
w/o $\mathcal{L}_\mathrm{sup}$                   & 74.5 & \multicolumn{1}{l|}{30.5/81.4/33.7} & 66.1      & 67.3/33.5     \\
w/o Variance              & 75.7 & \multicolumn{1}{l|}{31.2/81.0/30.5} & 66.9      & 70.9/33.8     \\
w/o Record                & 75.0 & \multicolumn{1}{l|}{31.3/86.3/36.1} & 66.4      & 69.9/36.4     \\ \bottomrule
\end{tabular}
\label{tab:ablative}
\vspace{-0.3cm}
\end{table}

We demonstrate the results in Table~\ref{tab:ablative} and summarize the findings as follows. (1) w/ SP achieves comparable fairness metrics compared to our full version, but its detection performance is inferior. We attribute this to the fact that static partition struggles to find genuinely informative invariant features that enhance performance, despite its ability to enforce fairness with sensitive labels. The confounding effects in rumor detection contexts are highly elusive, and limiting consideration to only a few of them prevents the identification of invariant features for undiscovered partitions. (2) Replacing the supervised contrastive loss with typical cross-entropy loss during the unfair partition stage results in poor performance. This implies that supervised contrastive loss is more effective than cross-entropy loss in discovering variant feature patterns caused by confounding sensitive attributes. Additionally, using cross-entropy loss in this context renders the first term in Eq. 2 meaningless, as the overall cross-entropy loss remains static across any partitions. Similarly, dropping the variance loss hinders the model's capability to learn generalizable invariant features that work for different groups. (3) The record of partition matrices is also necessary, enabling us to continually handle confounding effects and unfair causes, thereby enhancing the flexibility and dynamics of our method.

\subsubsection{Visual Evidence and Effective Intervention}
\begin{figure}[t]
\centering
  \begin{subfigure}{0.45\linewidth}
    \includegraphics[width=\linewidth]{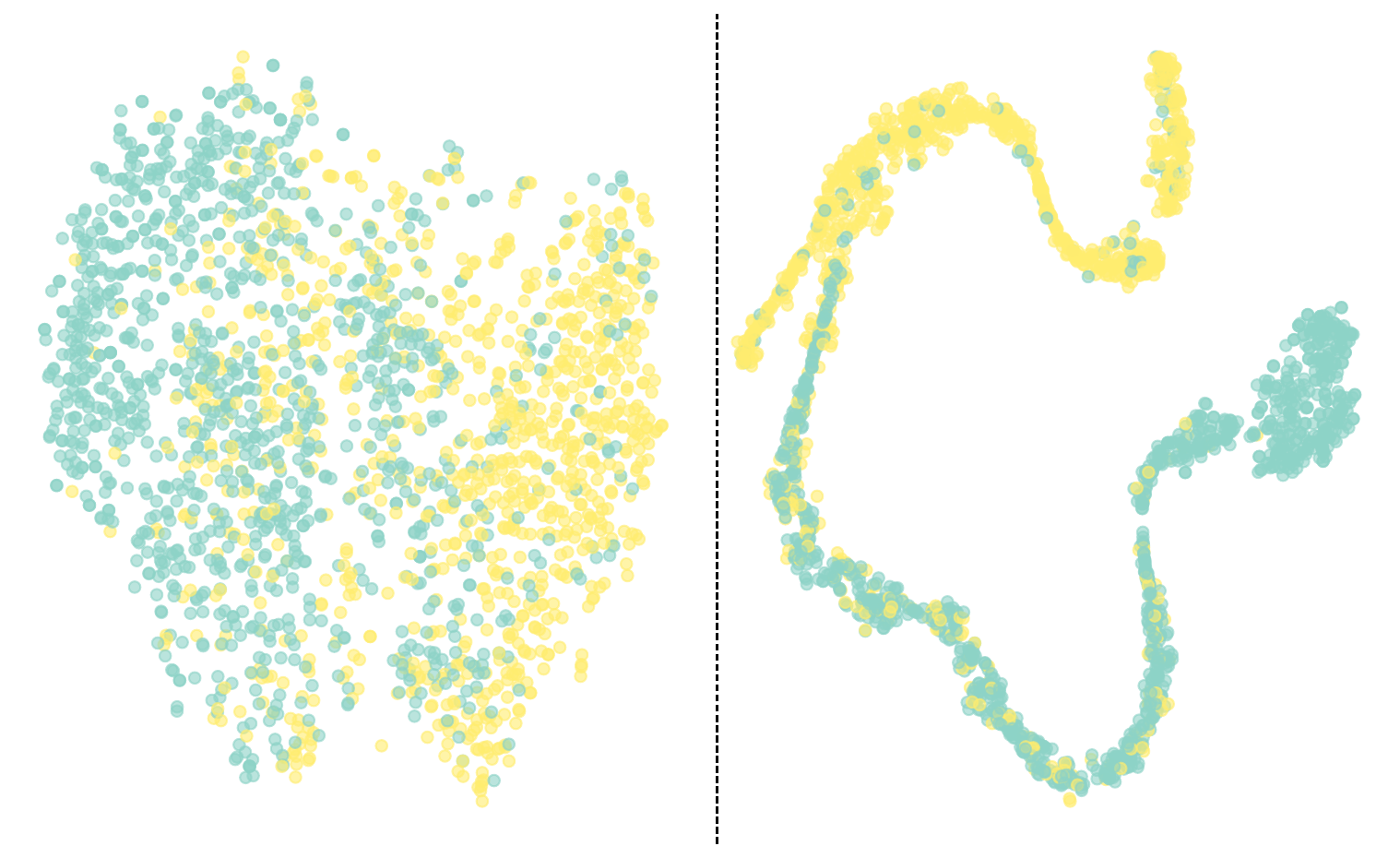}
    \caption{Ground true rumor labels}
  \end{subfigure}
  \hspace{0.4cm}
  \begin{subfigure}{0.45\linewidth}
    \includegraphics[width=\linewidth]{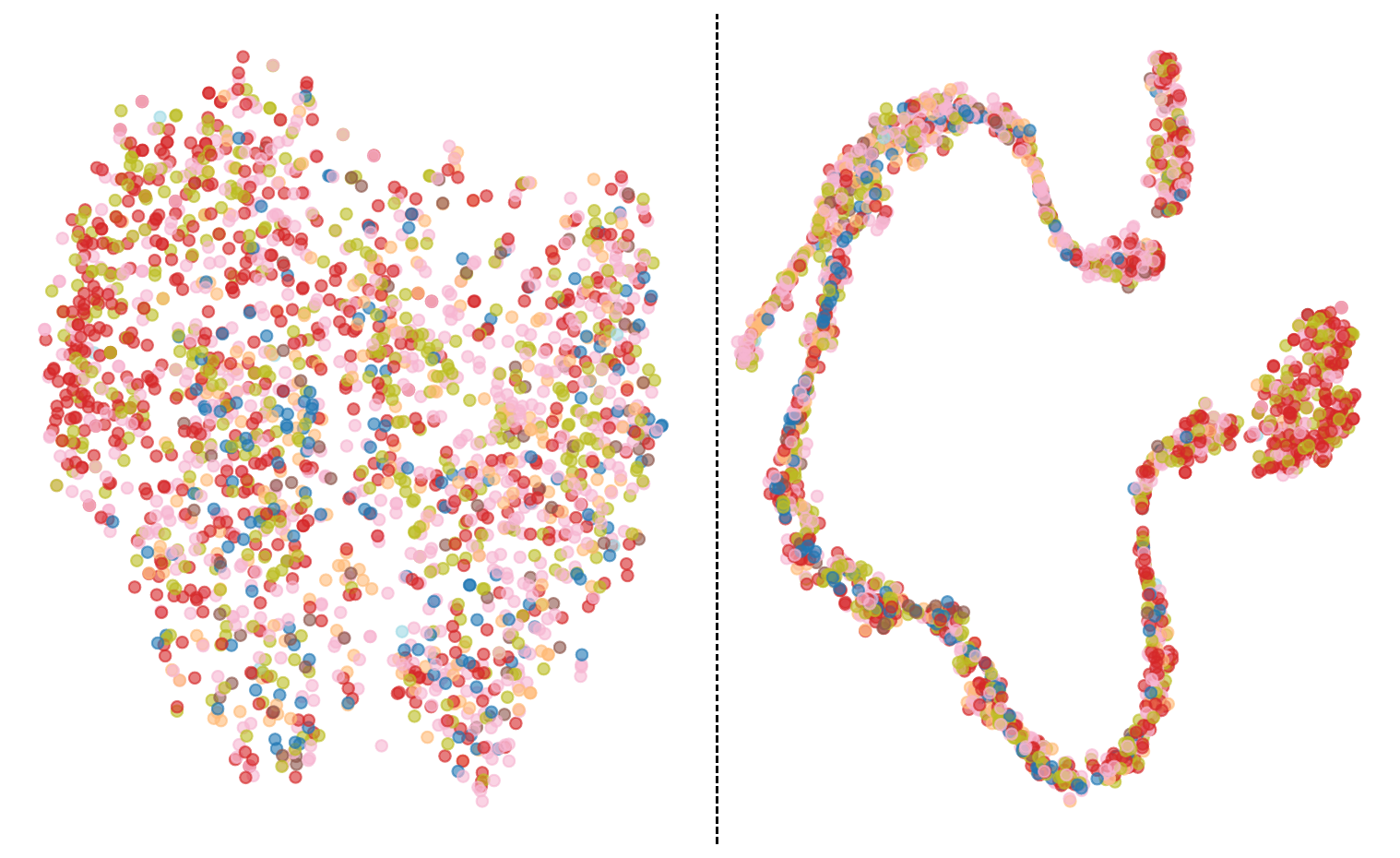}
    \caption{Topics}
  \end{subfigure}
    \begin{subfigure}{0.45\linewidth}
    \includegraphics[width=\linewidth]{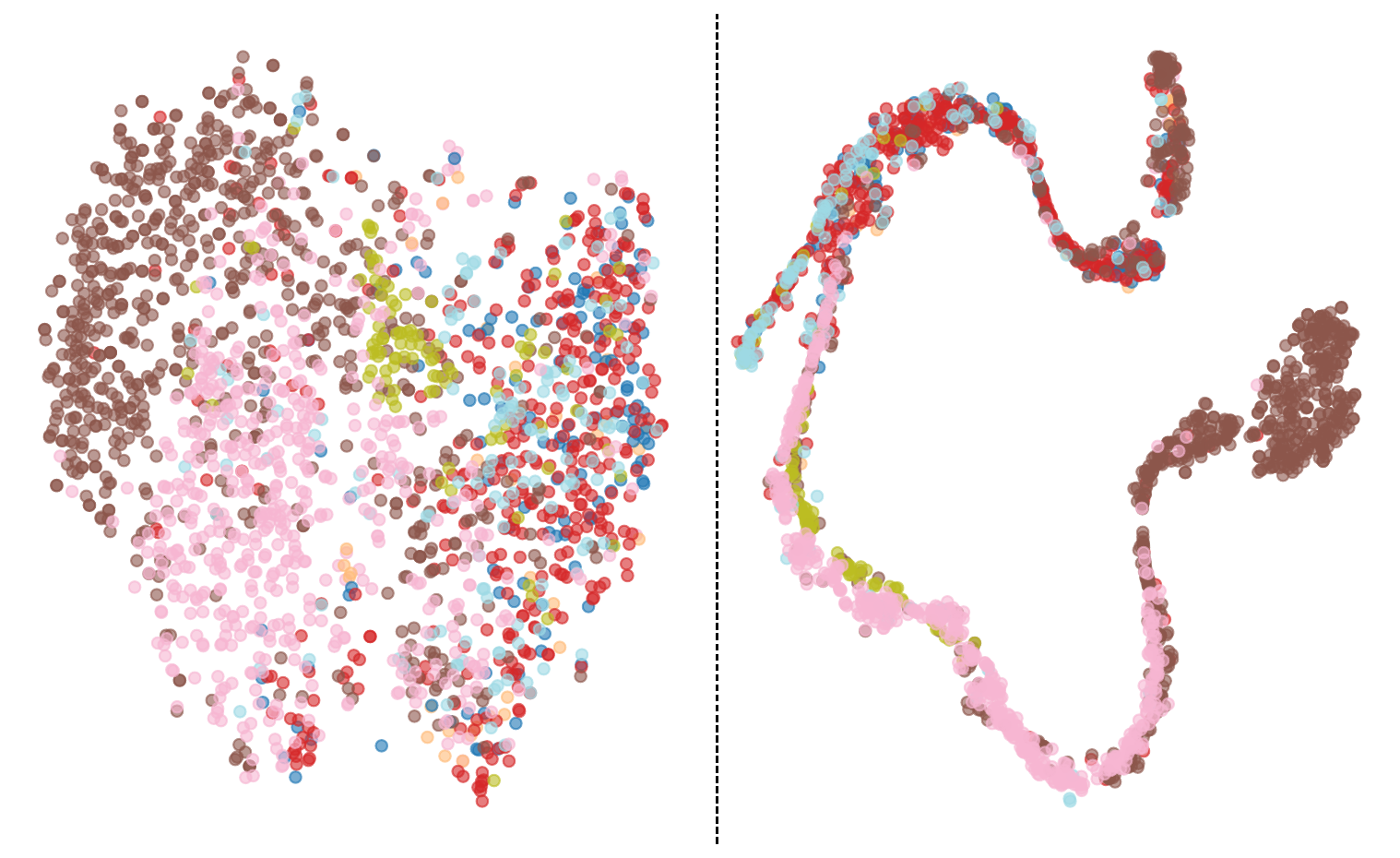}
    \caption{Platforms}
  \end{subfigure}
  \hspace{0.4cm}
    \begin{subfigure}{0.45\linewidth}
    \includegraphics[width=\linewidth]{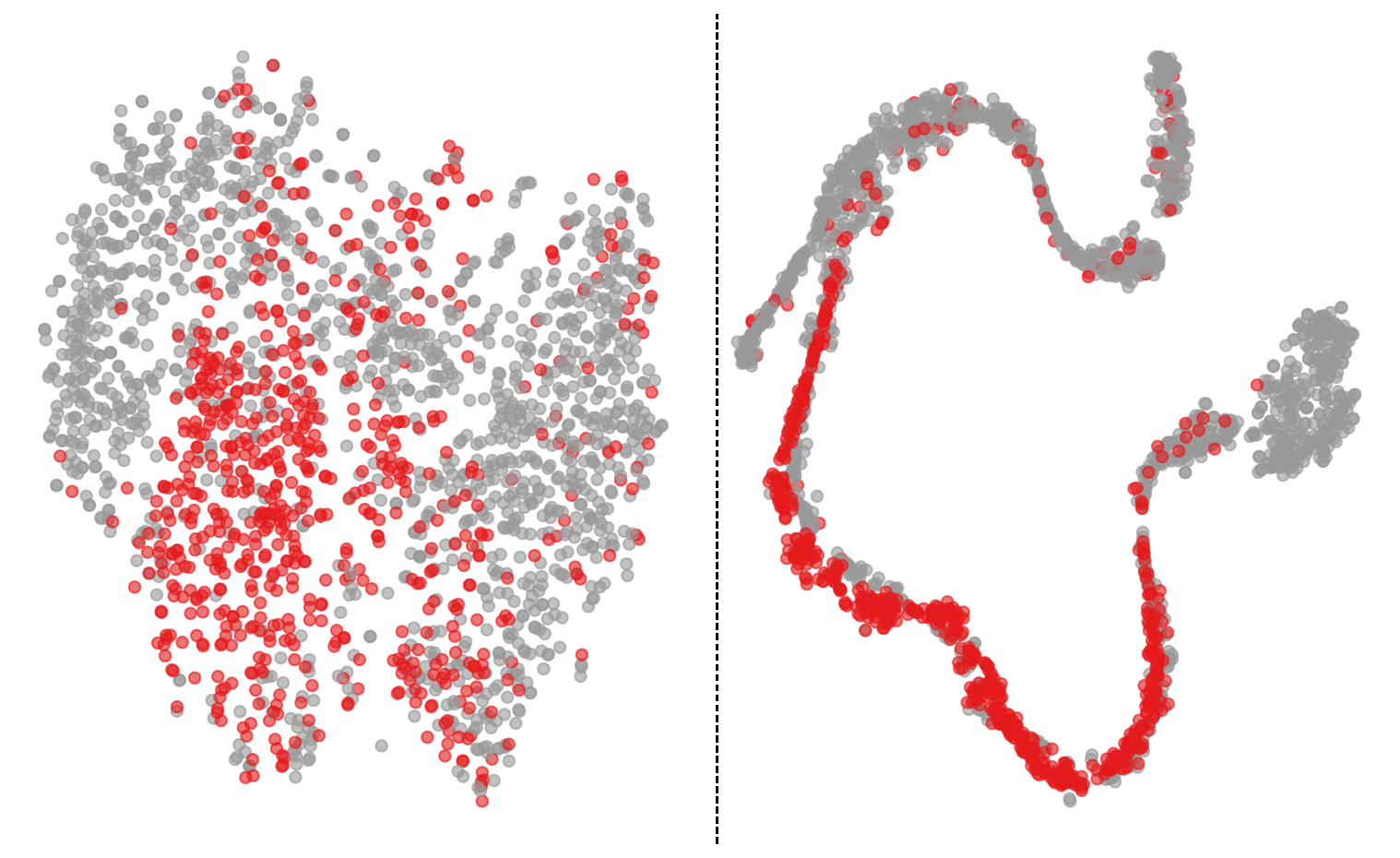}
    \caption{Authors}
  \end{subfigure}
\caption{Visualization of learned features. Left: MDFEND. Right: MDFEND w/ our method. Colors denote different groups. Fewer points appear visible with our method due to overlap.}
  \label{fig:visualization}
  \vspace{-0.5cm}
\end{figure}
To more thoroughly demonstrate that our method can improve the rumor detector's performance while addressing unfairness issues caused by confounding sensitive attributes, we visually present the learned features from our method and the best-performing method regarding detection performance, MDFEND. We color-code the points according to ground truth rumor/non-rumor labels, domain labels, platform labels, and certification status of authors from the \textit{FineFake} test set.

As demonstrated in Fig. \ref{fig:visualization}, the visualized learned representations from our method, compared to those from MDFEND, demonstrate: (1) discriminative power in identifying rumors and non-rumors, and (2) that the original learned representations are vulnerable to confounding sensitive attributes. In many regions, the visualized points originate from specific groups, and distributions of different groups are scattered, making it easy to learn shortcuts from them. In contrast, with our method, the points from different groups are uniformly mixed on the manifold, tightly clustered, and even feature patterns from the same groups are diverse, preventing the model from making biased choices.  The consistent observations across subfigures further confirm that our method effectively considers fairness for different sets of groups during training.

\begin{figure}[t]
    \centering
    \includegraphics[width=\linewidth]{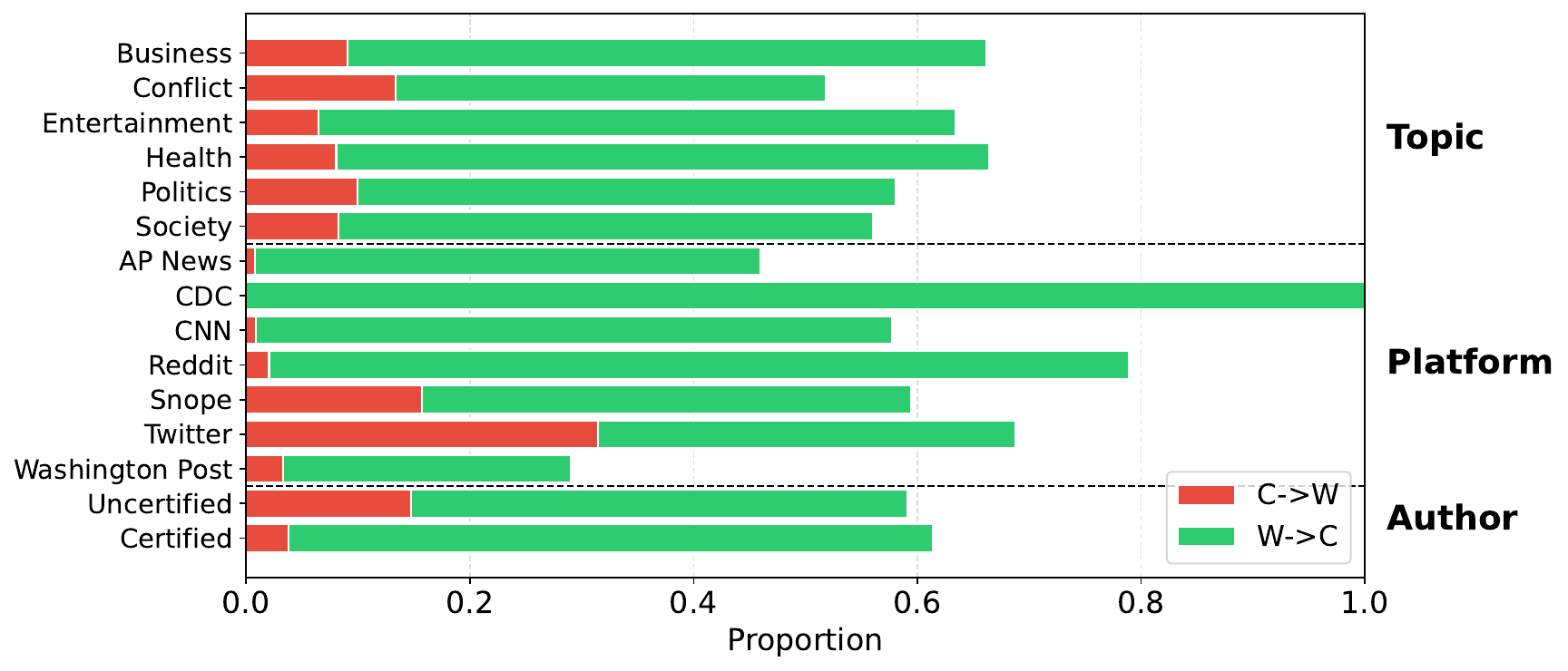}
\caption{Intervention effects of our method on the base model. W->C: proportion of base model's wrong predictions corrected by our method; C->W: proportion of correct predictions made incorrect.}
    \label{fig:intervention}
    \vspace{-0.2cm}
\end{figure}

To examine how our method conducts effective intervention on the base model regarding different groups, we provide further evidence in Fig. \ref{fig:intervention}. It is evident that in most cases where the base model makes incorrect predictions, our method helps correct them, while there is minimal chance that our method will mislead the base model's already correct predictions.
\section{Conclusion}
We propose a method enhancing both detection performance and fairness across groups in rumor detection. Our key findings reveal that: (1) multiple overlooked sensitive attributes confound existing detectors, (2) addressing these attributes improves detection accuracy and fairness, but (3) focusing on specific attributes can disadvantage groups defined by unconsidered ones. Future work will establish a fair rumor detection benchmark and explore sparse sensitive attribute integration for improved effectiveness.
\vspace{-0.3cm}
\bibliographystyle{ACM-Reference-Format}
\bibliography{sample-base}

\end{document}